\documentclass[a4paper,12pt]{article}
\usepackage[dvips]{graphicx}

\usepackage{amssymb}
\usepackage{amsmath}
\usepackage{longtable}

\title{Magnetic Monopole Bibliography-II}
\author{S. Balestra$^{1}$, G. Giacomelli$^{1,2}$, M. Giorgini$^{3}$, L. Patrizii$^{1}$, V. Popa$^{4}$, \\ Z. Sahnoun$^{1,5}$ and V. Togo$^{1}$}
\date{\small $^{1}$ INFN sez. Bologna, v.le Berti Pichat 6/2, I-40127 Bologna, Italy.\\
$^{2}$ Phys. Dept.,  Univ. Bologna, v.le Berti Pichat 6/2, I-40127 Bologna, Italy.\\
$^{3}$ INAF - IASF Milano, via E. Bassini 15, I-20133 Milano, Italy.\\
$^{4}$ Institute of Space Sciences, Bucharest R-76900, Romania.\\
$^{5}$ Astrophys. Dept. CRAAG, B.P. 63, Bouzareah, Algiers, Algeria.}

\begin{document}
\maketitle

\begin{abstract}
The bibliography compilation on magnetic monopoles is updated to include references from 2000 till mid 2011. It is intended to contain all experimental papers on the subject and only the theoretical papers which have specific experimental implications.
\end{abstract}

\section*{Introduction}\label{aba:sec1}
Magnetic Monopoles (MMs) were introduced in 1931 by Dirac in order to explain the 
quantization of the electric charge~\cite{dirac}. He established the relation between 
the elementary electric charge $e$ and a basic magnetic charge $g$:
	$e \,  g=n\hbar c/2= ng_{D}$,
where $n$ is an integer, $n=1,2,...$, $g_{D}=\hbar c/2e = 68.5 e$ is the unit Dirac charge.

The existence of magnetic charges and of magnetic currents would symmetrize in form  Maxwell's equations, but the symmetry would not be perfect since numerically $g>>e$ in the cgs symmetric system of units:

\begin{equation}
\begin{array}{ll}
            \nabla \cdot \vec{E} = 4 \pi \rho_{e} & \qquad \nabla \times \vec{B} = \frac{\displaystyle{4 \pi}}{\displaystyle{c}} \vec{j_{e}} + \frac{\displaystyle{\partial\vec{E}}}{\displaystyle{\partial t}} \\ [10pt]

            \nabla \cdot \vec{B} = 4 \pi \rho_{m} & \qquad \nabla \times \vec{E} = \frac{\displaystyle{4 \pi}}{\displaystyle{c}} \vec{j_{m}} - \frac{\displaystyle{\partial\vec{B}}}{\displaystyle{\partial t}} \\
\end{array}
\end{equation}

\noindent If the couplings are energy dependent they could converge to a single common value at very high energies \cite{derujula}.\\ 

\indent There was no prediction for the classical MM mass. A rough estimate, obtained assuming that the classical monopole radius is equal to the classical electron radius yields $m_M \simeq \frac{\displaystyle{g^{2}m_e}}{\displaystyle{e^{2}}} \simeq n \ 4700\  m_e \simeq n \ 2.4\  GeV/c^{2}$. This mass could be considerably larger if the basic charge is $e/3$ and $n > 1$.\\

Later, in 1974, t'Hooft and Polyakov \cite{thooft, polyakov, craigie} demonstrated that MMs are natural solutions in Grand Unified Theories (GUT) of the strong and electroweak interaction. The GUT unified group eventually breaks spontaneously into subgroups, one of which could be the electromagnetic U(1) subgroup. For instance starting with the SU(5) GUT group :
\begin{equation}
\footnotesize
    \begin{array}{ccccc}
        {} & 10^{15}\ GeV & {} & 10^{2}\ GeV & {} \\
        SU(5) & \longrightarrow & SU(3)_{C}\times \left[ SU(2)_{L}\times U(1)_{Y}\right] & \longrightarrow & SU(3)_{C}\times U(1)_{EM} \\
       {} & \small10^{-35}s & {} & \small10^{-9}s & {}
    \end{array}
\end{equation}
The MM mass is related to the mass of the X, Y carriers of the unified interaction, $ m_{M}\ge m_{X}/G$, where G is the dimensionless unified coupling constant at energies E $\simeq m_{X}$. If $m_{X}\simeq 10^{14}-10^{15}$ GeV, $G\simeq0.025$, and the breaking yields immediatly a U(1) group, then $m_{M}>10^{16}-10^{17}$ GeV. Heavier MM masses are expected if gravity is brought into the unification picture, and in some SuperSymmetric models. GUT MMs could only have been produced in the very first instant of the Early Universe.\\

If the U(1) group would appear in a later phase transition, this could have lead to the production of Intermediate Mass Monopoles (IMMs)~\cite{lazaride, kephart} which could be multiply charged and have masses   m$_{M}$ $\sim 10^{7} \div 10^{13}$ GeV. For example starting with the SO(10) GUT group one could have:

\begin{equation}
\footnotesize
    \begin{array}{ccccc}
        {} & 10^{15}\ GeV & {} & 10^{9}\ GeV & {} \\
        SO(10) & \longrightarrow & SU(4)\times SU(2)\times SU(2) & \longrightarrow & SU(3)\times SU(2)\times U(1) \\
       {} & \small10^{-35}s & {} & \small10^{-23}s & {}
    \end{array}
\end{equation}

These MMs may be accelerated to relativistic velocities in one galactic magnetic field domain. It was hypothesized that very energetic IMMs could yield the highest energy cosmic rays~\cite{bhatta}.\\
 \indent Electrically charged Monopoles, so-called Dyons, may arise as quantum mechanical excitations or as M-p, M-nucleus composites (Monopole-Dipole interaction).\\
\indent Searches for Dirac Magnetic monopoles were mainly carried out at accelerators, at colliders in $e^{+}e^{-}$, $e^{+}p$, $p \overline{p}$ and $p p$ collisions. GUT and Intermediate Mass MMs, given their large expected masses, can only be searched for as relic particles from the Early Universe in the cosmic radiation.\\

The main signature for MM detection is that they should be highly ionizing. In fact, a fast MM with magnetic charge $g_{D}$ and velocity $v=\beta c$ behaves like an equivalent electric charge $(ze)_{eq}=g_{D}\beta$ losing energy mainly by ionization; thus for $\beta>10^{-1}$, the energy loss of a $g_{D}$ MM is $\sim (68.5)^2 \sim 4700$ times that of a Minimum Ionizing Particle.\\

Slow poles, ($10^{-4}<\beta<10^{-2}$) lose energy by ionization or excitation of atoms and molecules of the medium (``electronic'' energy loss) or by yielding kinetic energy to recoiling atoms or nuclei (``atomic'' or ``nuclear'' energy loss). Electronic energy loss predominates for $\beta>10^{-3}$. In noble gases and for monopoles with $10^{-4}<\beta<10^{-3}$ there is an additional energy loss due to atomic energy level mixing and crossing (Drell effect). \\

\begin{figure}[h]
	\begin{center}
		\includegraphics[width=0.8\textwidth]{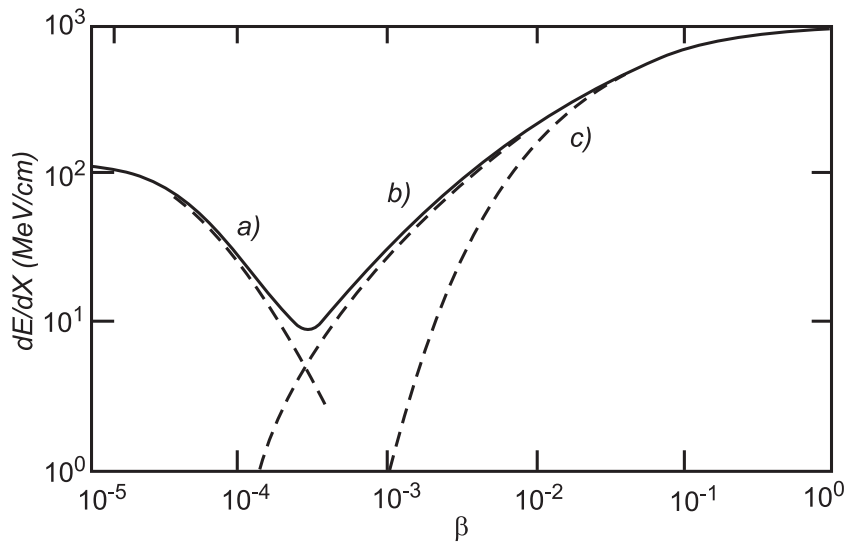}
	\end{center}
	\caption{The energy losses, in MeV/cm, of $g=g_{D}$ MMs in liquid hydrogen vs ${\beta}$. Curve a) corresponds to elastic monopole--hydrogen atom scattering; curve b) to interactions with energy level crossings; curve c) describes the ionization energy loss.}
	\label{fig:perdita-di-energia}
\end{figure}

\indent At very low velocities ($v<10^{-4}c$), MMs may lose energy in elastic collisions with atoms or with nuclei. The energy is released to the medium in the form of elastic vibrations and/or infrared radiation~\cite{derkaoui1}.\\
\indent In Fig. \ref{fig:perdita-di-energia} are shown the main different energy loss mechanisms at work in liquid hydrogen of a $g=g_{D}$ MM versus $\beta$~\cite{gg+lp}.

According to G. Lochak \cite {lochak} the Dirac equation for massless particles would lead to a massless leptonic magnetic monopole which interacts also weakly. These monopoles may be considered as magnetically excited neutrinos, and could be produced by electromagnetic impulses leading to nuclear trasmutation. 
For beta radioactive nuclei, e.m. pulses could shorten their lifetimes with the emission, in a magnetic field, of monopoles instead of neutrinos. These statements are interesting, but need more solid experimental cross checks.\\

\indent There are also discussions in the condensed matter field about magnetic monopoles as quasiparticles in Spin-Ice; classical analogues of these particles occurs as excitations of the topological ground state \cite{spinIce1,spinIce2,spinIce3,spinIce4}, ...\\

\indent Since the publications of t'Hooft and Polyakov \cite{thooft, polyakov, craigie} a very large number of theoretical studies were made on magnetic monopoles, which are very interesting mathematical entities. There still are many theoretical works on Dirac MMs, on Intermediate mass and GUT Monopoles. There are also many papers concerning possible monopoles bound in (hiding in) $M\overline{M}$ states, as Dyons and on very light monopoles \cite{vento}.\\

\indent The present paper gives a bibliography of most recent publications on monopoles recalling also the most significant early papers; it is an update of the 2000 magnetic monopole bibliography \cite{giacomelli}. This bibliography is intended to contain all the experimental papers from year 2000 to may 2011 on the subject and only the main theoretical papers and those which have specific experimental implications.\\

\indent Experimental direct and indirect searches for classical Dirac monopoles at accelerators and colliders and searches for MMs in the cosmic radiation were and still are performed.
By direct searches we intend detecting monopoles immediately after their production in high-energy collisions, or in the cosmic radiation. Indirect searches for MMs include a variety of searches in matter as for example, when analysing a piece of matter which was exposed at an accelerator beam or at cosmic rays. Large neutrino telescopes also performed searches for very fast, relativistic, Intermediate mass MMs using the cherenkov light emitted in water or ice \cite{baikal1, antares, baikal, amanda, amanda2}.\\

\indent In the Particle Data Books have been reported regularly the magnetic monopoles searches, for 2010 see \cite{pdg}, and also some review papers, for 2010 see \cite{pdgmms}.\\

\indent As a by-product of MM searches some experiments were able to set upper limits on the flux of some other massive exotic particles such as nuclearites and Q-balls. A few theoretical works and all experimental results are also given in this bibliography \cite{derujulaN, halzen, pacheco, kusenko, kusenko2, madsen, bakari, popa, kumar, cecchiniN, popaN, amandaN, skq, cecchiniN2, giacomelliN3, giacomelliN3, sahnoun, pavalas}.\\

\begin{figure}[h]
	\begin{center}
		\includegraphics[width=0.81\textwidth]{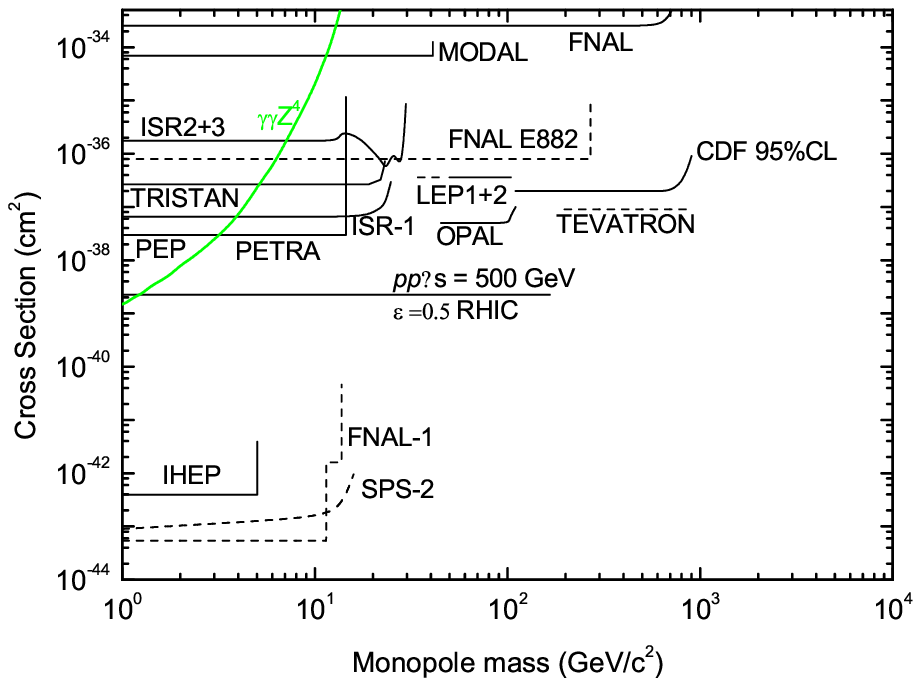}
	\end{center}
	\caption{Compilation of the cross section upper limits on classical Dirac Monopole produced at fixed target accelerators and at colliders. Solid lines are for direct experiments and dashed ones are for indirect experiments. }
	\label{fig:MMUpLim}
\end{figure}

\begin{figure}[h]
	\begin{center}
		\includegraphics[width=0.8\textwidth]{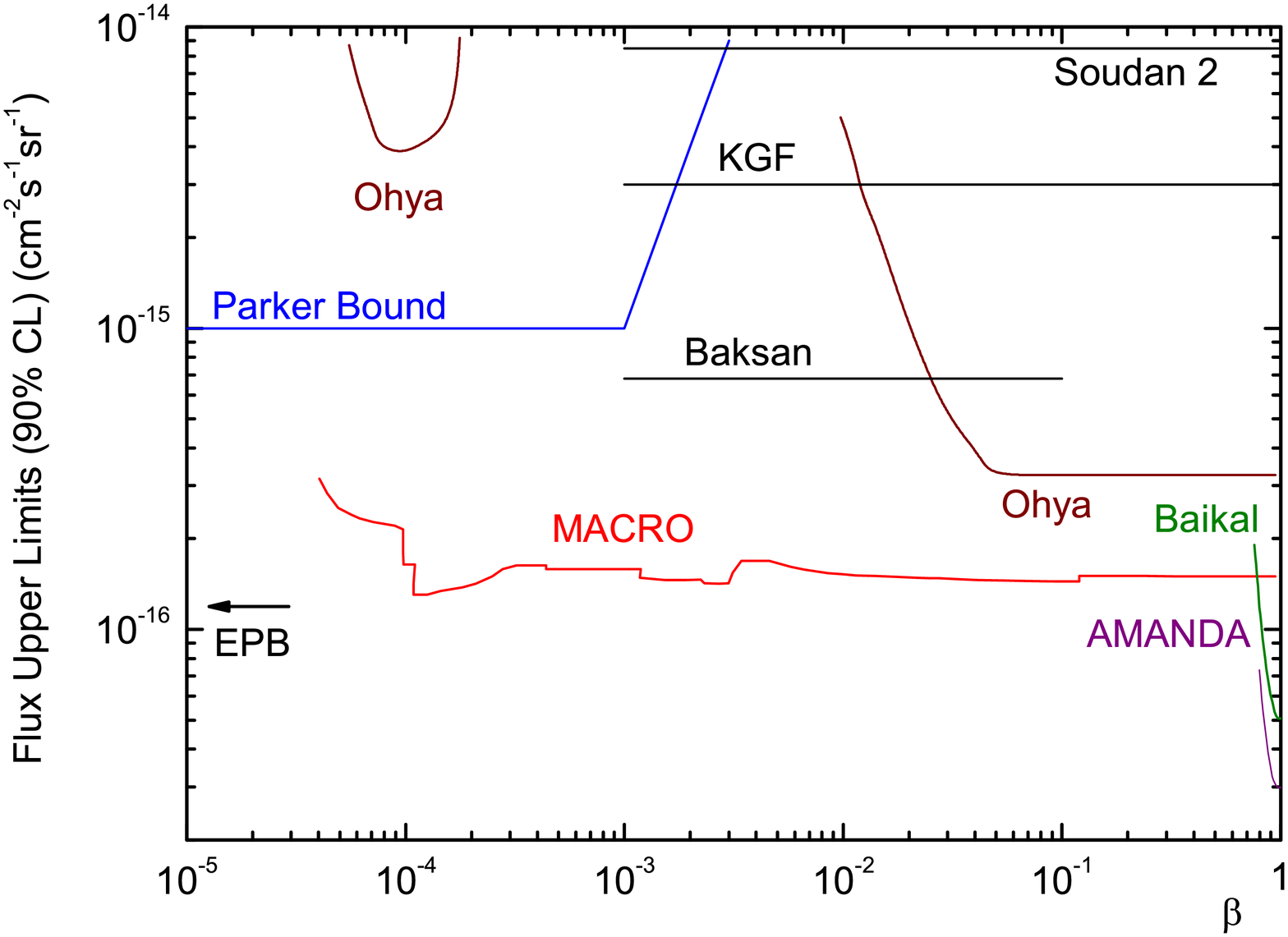}
	\end{center}
	\caption{The 90\% CL direct upper limits vs $\beta$ for GUT magnetic monopoles with $g=g_{D}$ in the penetrating CR.}
	\label{fig:GUTlimit}
\end{figure}

\begin{figure*}
	\begin{center}
    \resizebox{!}{5.4cm}{\includegraphics{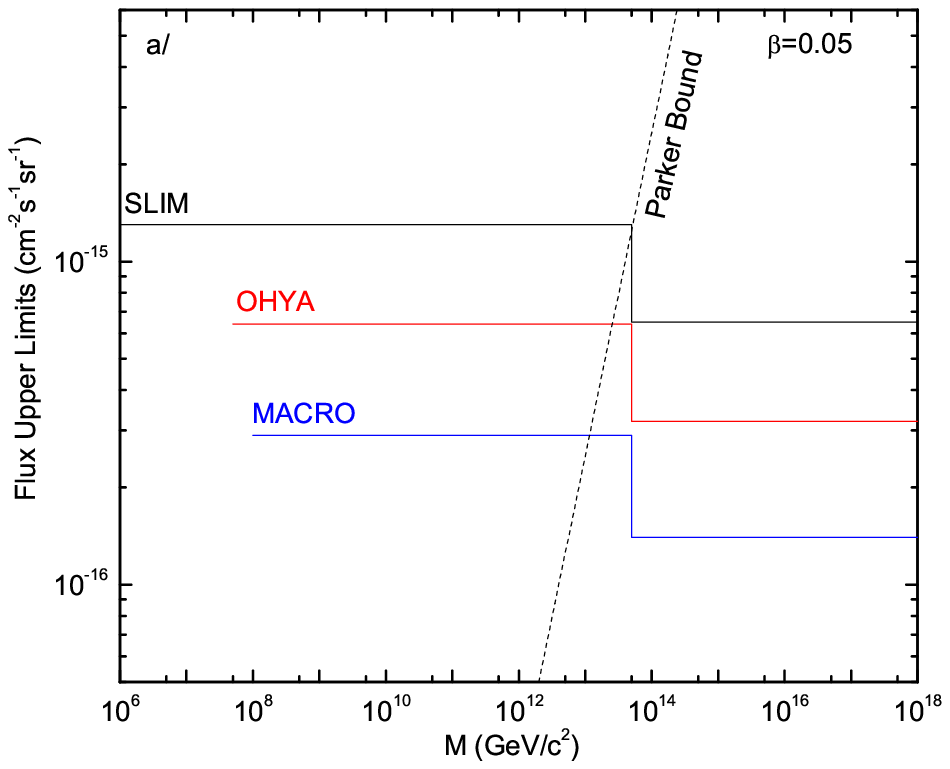}}
    \hspace{0.0cm}
    \resizebox{!}{5.5cm}{\includegraphics{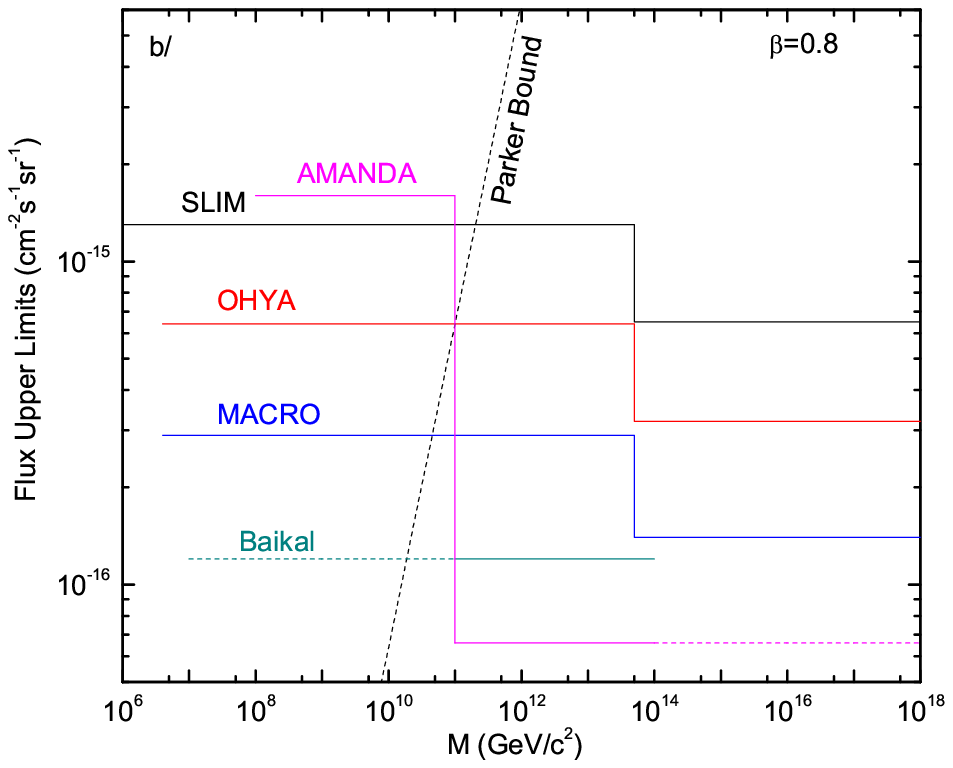}}
	\caption{Experimental 90\% CL flux upper limits versus mass for MMs with: a) $\beta \sim 0.05$, b) $\beta \sim 0.8$ at the detector level, as given from different experiments.}
	\label{fig:monopoles}
	\end{center}
\end{figure*}

\begin{figure}
	\begin{center}
\resizebox{12cm}{7.8cm}{\includegraphics{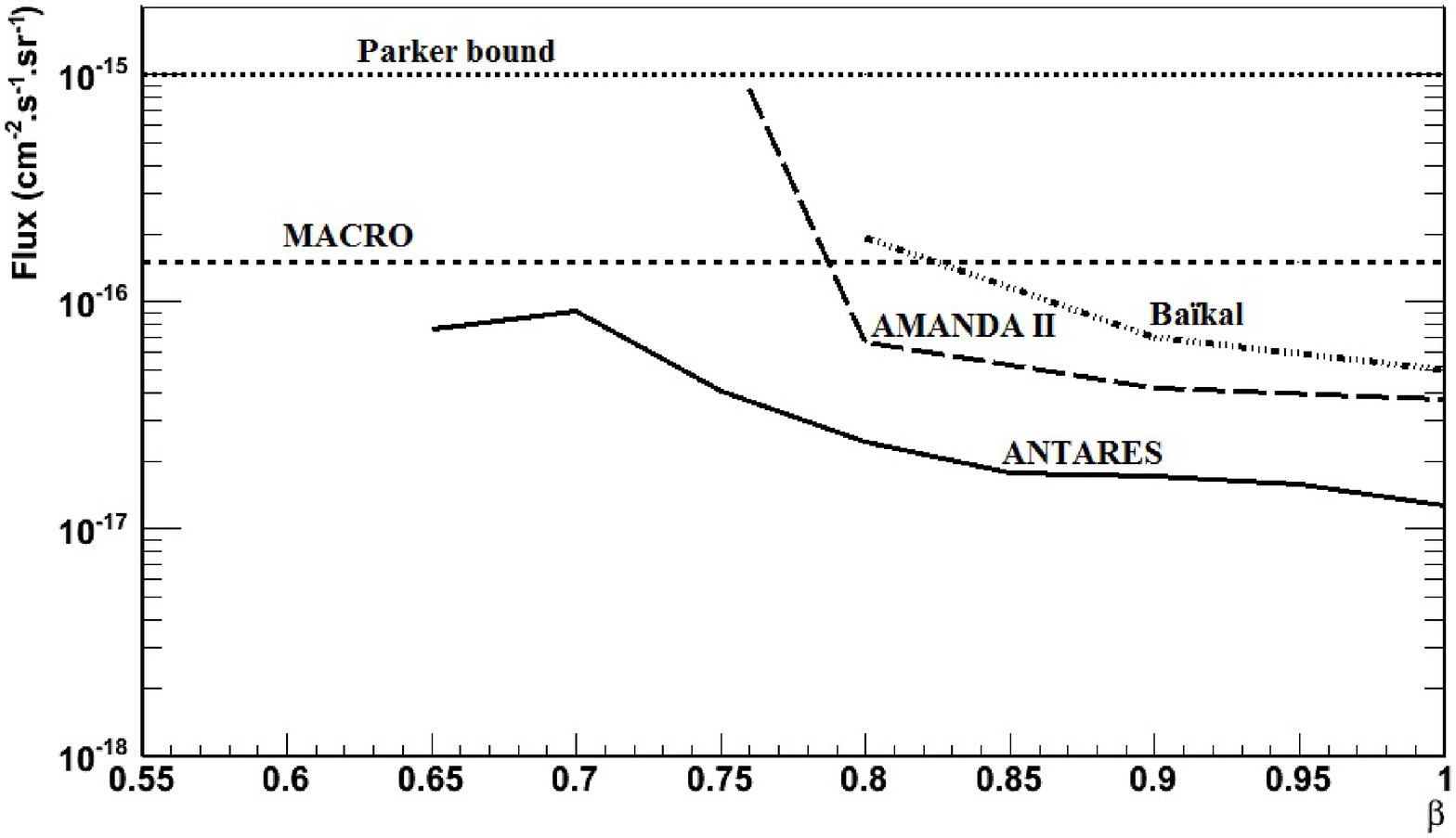}}
	\end{center}
	\caption{90\% CL upper limits on fast intermediate mass MMs in the cosmic radiation. The limits were obtained by large underground detectors (MACRO) and very large neutrino telescopes (AMANDA, BAIKAL, ANTARES). The dotted horizontal line at $10^{-15}$ cm$^{-2}$s$^{-1}$sr$^{-1}$ is the Parker phenomenological limit.}
	\label{fig:antaresflux}
\end{figure}

\begin{figure}[h]
	\begin{center}
		\includegraphics[width=0.75\textwidth]{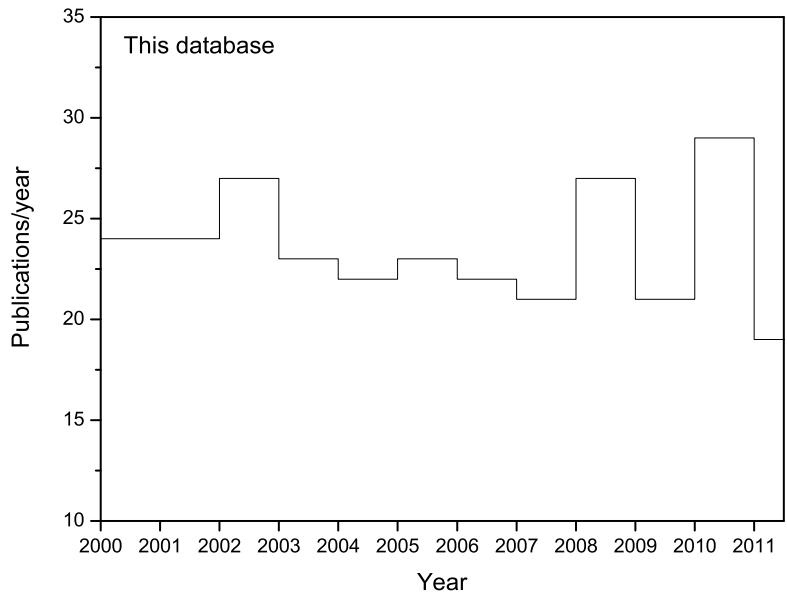}
	\end{center}
	\caption{Histogram of the yearly number of papers cited in the present database.}
	\label{fig:ThisDatabase}
\end{figure}

\begin{figure}[h]
	\begin{center}
		\includegraphics[width=0.75\textwidth]{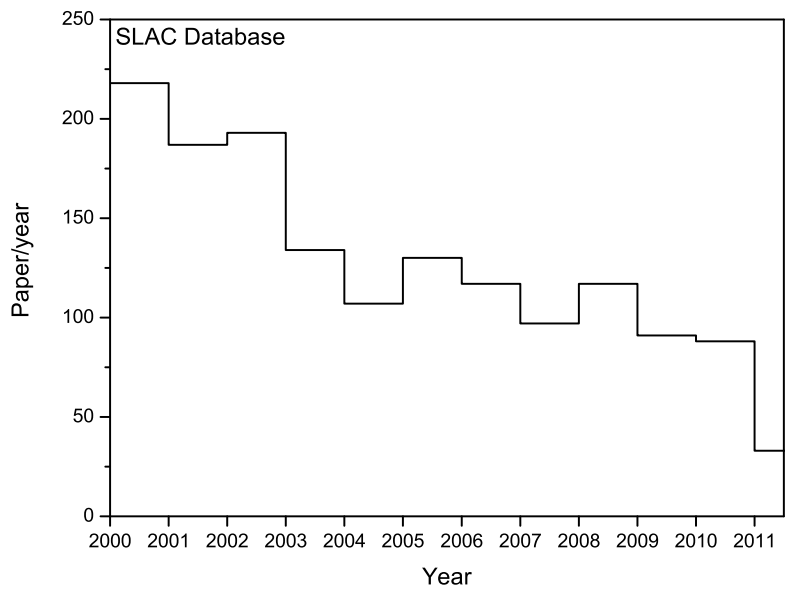}
	\end{center}
	\caption{Histogram of the yearly total number of papers on MMs (experimental, theoretical, phenomenological), published between 2000 and May 2011, as given by the SPIRES-SLAC database.}
	\label{fig:SLACDatabase}
\end{figure}

\indent The experimental limits obtained for the cross section production of Dirac monopoles at accelerators are shown in Fig.\ref{fig:MMUpLim}. The 90\%  CL flux upper limits set by searches for MMs in the cosmic radiation are given in Figs. \ref{fig:GUTlimit}, \ref{fig:monopoles} and \ref{fig:antaresflux}. Figs. \ref{fig:ThisDatabase} and \ref{fig:SLACDatabase} show histograms of the yearly number of papers on MMs cited in the present database and given in the SPIRES-SLAC database (papers considered contained ``magnetic monopole'', ``monopoles'' or ``dyon'' in the title or ``magnetic monopole'' as a keyword). Note that there is an important difference between the number of papers in this database and the SLAC database which at least implies that most papers are theoretical.

\section{Acknowledgments}
We acknowledge the collaboration from many colleagues. Z. S.  thanks INFN, Sez. Bologna for providing FAI Grants for foreigners.

\end{document}